\documentclass[12pt]{article}
\usepackage[top=2cm, bottom=2cm, left=2cm , right=2cm]{geometry}
\usepackage{cite}
\usepackage{amsmath,amssymb,amsfonts}
\usepackage{algorithmic}
\usepackage{graphicx}
\usepackage{textcomp}
\usepackage{xcolor}

\usepackage{wrapfig}
\usepackage{booktabs} 
\usepackage{tikz}
\usepackage{framed}
\usetikzlibrary{arrows.meta}
\usepackage{caption}
\usepackage{lipsum}
\usepackage{color}
\usepackage{ifthen}
\usepackage{array}
\usepackage{authblk}

\newtheorem{definition}{Definition}
\newtheorem{theorem}{Theorem}
\newtheorem{lemma}{Lemma}
\newtheorem{corollary}{Corollary}
\newcommand{\qed}{$\Box$~}

\def\BibTeX{{\rm B\kern-.05em{\sc i\kern-.025em b}\kern-.08em
    T\kern-.1667em\lower.7ex\hbox{E}\kern-.125em}}

\begin{document}

\title{Lightweight FEC: Rectangular Codes with Minimum Feedback Information
\thanks{For financial support, we are grateful to:
  \textit{Thales Communications \& Security}, project TCS.DJ.2015-432;
  \textit{Agence Nationale de la Recherche Technique}, project 2016.0097.}
}

\author{
{Binh-Minh Bui-Xuan\break
\textit{LIP6 -- CNRS -- Sorbonne Universit\'e, Paris, France.}\break
\textit{buixuan@lip6.fr}
}
\and
{Pierre Meyer\break
\textit{\'Ecole Normale Sup\'erieure de Lyon, Lyon, France.}\break
\textit{pierre.meyer@ens-lyon.fr}
}
\and
{Antoine Roux\break
\textit{LIP6 -- CNRS -- Sorbonne Universit\'e, Paris, France.}\break
\textit{Thales Communications \& Security, Genevilliers, France.}\break
\textit{antoine.roux@lip6.fr}
}
}

\maketitle

\begin{abstract}
We propose a hybrid protocol combining a rectangular error-correcting code - paired with an error-detecting code - and a backward error correction in order to send packages of information over a noisy channel.
We depict a linear-time algorithm the receiver can use to determine the minimum amount of information to be requested from the sender in order to repair all transmission errors.
Repairs may possibly occur over several cycles of emissions and requests.
We show that the expected bandwidth use on the backward channel by our protocol is asymptotically small.
In most configurations we give the explicit asymptotic expansion for said expectation.
This is obtained by linking our problem to a well known algorithmic problem on a gadget graph, feedback edge set.
The little use of the backward channel makes our protocol suitable where one could otherwise simply use backward error correction, e.g.\ TCP, but where overly using the backward channel is undesirable.
We confront our protocol to numerical analysis versus TCP protocol.
In most cases our protocol allows to reduce the number of iterations down to 60\%, while requiring only negligibly more packages.

\end{abstract}



\clearpage
\section*{Introduction}
\label{sec:introduction}

Traditionally, \emph{forward error correction} (FEC) is used to correct a limited amount of transmission errors over unidirectional channels, for example from an untrusted to a trusted network~\cite{B2010}.
If the unidirectionality constraint can be relaxed, a small backward channel can allow the receiver to request additional information in order to correct more or even all of the transmission errors. This relaxation is even necessary if we wish to pass encrypted data, as in that case no amount of corruption, however small, is tolerable. However, for security reasons, it is important that use of this channel be confined to an absolute minimum.

Error correction over a noisy channel can also be achieved by \emph{automatic repeat request} (ARQ).
ARQ protocols require an error detector such as the \emph{user datagram protocol} (UDP) and are based on an acknowledgement system: the sender re-emits each packet until the receiver acknowledges it has been received without errors. The main idea behind forward error correction is to use an error-correcting code to send redundant information over the channel in the hope that transmission errors will leave enough to retrieve the original message. Purely information re-emission based protocols can lead to a prohibitive number of cycles of exchanges between sender and receiver, especially if the \emph{round-trip time}, a.k.a. the ping, is high. On the other hand, no protocol based solely on an error-correcting code can guarantee a reliable transmission: there is always a non-zero probability, growing with the channel's bit error rate, that the message is not transmitted in full. Note that if the message is encrypted, then partial information is useless. Moreover, the redundant information can prove costly in bandwidth use. One solution is to combine backward and forward error correction in order to find a trade-off between transmission delay, success probability, and bandwidth use. \emph{Hybrid ARQ} (HARQ) is based on this principle, and shows good results on sufficiently noisy channels (\cite{centenaro},\cite{chelli},\cite{lin}).

In this paper, we propose a protocol combining a rectangular error-correcting code and backward error correction in order to send packets of information over an erasure channel.
These packets are authenticated with the UDP protocol. Note that there are other error-correcting codes which do not use as much redundancy (\cite{reed-solomon},\cite{chatzigeorgiou},\cite{gallager},\cite{berrou}), but our choice of rectangular codes is motivated by their structural simplicity. Indeed they can be seen as a system of equations, where the unknown variables correspond to the packets erroneously transmitted. It follows that a set of errors is repairable if and only if the unknown variables corresponding to erroneous packets can be eliminated one by one, i.e. the system of equations presents a form of acyclicity. We present a linear-time cycle-breaking algorithm the receiver can use to determine the minimum amount of information to be requested from the sender at the next iteration. This extra step reduces bandwidth use and the number of iterations.
Numerical experiments of our protocol show a significant reduction in the number of iterations, $\approx 60\%$, of what is needed by the \emph{Transmission Control Protocol} (TCP).
Further, they show that the amount of information that goes from the sender to the receiver is asymptotically the same with or without the rectangular code: the redundancy introduced by the code is compensated by the minimality of the feedback requests.
In a sense, our protocol condenses TCP into fewer iterations which suggests it could prove suited to transmissions in space, where round-trip time is high. Furthermore, computations are expensive for
deep-space probes making the simplicity of rectangular codes all the more interesting.

The paper is organised as follows.
[Section~\ref{sec:rectangular-codes}] defines the proposed protocol. [Section~\ref{sec:analysis}] studies two quality indicators of the protocol: the expected bandwidth use and the expected number of iterations, assuming transmission errors are distributed independently and with uniform probability. [Section~\ref{sec:results}] presents simulated experimental results and illustrates the results of [Section~\ref{sec:analysis}].

\section{Rectangular codes with feedback}
\label{sec:rectangular-codes}

We denote by $\mathbb{N}_n$ the integer interval $[\![0;n[\![$. 
In this section we describe our hybrid forward error correction scheme. Suppose that we have a message $\mathcal{M}$, which we fix throughout this section, that the sender wishes to send to the receiver. The message is divided as an ordered sequence of $K$ bit vectors, all of the same size, where $K$ is a parameter of the protocol. For our protocol to work, the receiver must be able to determine which of the bit vectors were corrupted during transmission. Each of them is authenticated using the UDP protocol, and in particular the CRC-32 algorithm. Throughout this paper, we call \emph{packet} such an authenticated bit vector, and denote by $(len)$ the common length of all packets of the message. In practice, it should be thought of as being of order at least $2^{10}$ -- that is, the packets are of the order of 1~kB. We assume no packets are falsely flagged as correct. In the following, $\oplus$ denotes the binary bitwise-xor operator on bit vectors. When we say that we `sum' -or `xor'- two packets, it should be understood that we perform a bitwise sum modulo two of the vectors of information embedded in the packets, dismissing the authentication bits, and then authenticate this new bit vector to obtain a packet. On a side note, the UDP protocol adds an index to each packet, which means packets in a single block can safely be received in any order.

\subsection{Rectangular codes}

A \emph{linear block code} is an error-correcting code which acts on $K$ bits of input data to produce $N$ bits of output data $(N,K)$ via a linear transformation. The input and output of the codes we consider here are not single bits but vectors of fixed length $(len)$ called \emph{packets}, so strictly speaking they are $(N*(len),K*(len))$ block codes. However, for simplicity we shall say that they act on $K$ packets to produce $N$ packets. Further, since the input data is embedded in the encoded output they are \emph{systematic codes}. We will restrict our study to two-dimensional rectangular codes and we will therefore, for simplicity's sake, only define rectangular codes in two dimensions.

\subsubsection*{Encoding Scheme}

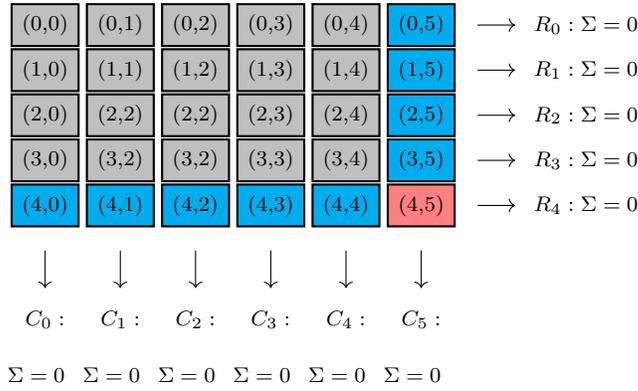
\begin{figure}
  \begin{center}
  \begin{tabular}{c c}
    packet $\overrightarrow{C}_k$ & Coordinates\\
    \toprule
    $0\leq k <K$ & $\phi(k)$\\
    \midrule
    $K\leq k <K+n$ & $(k-K,m)$\\
    \midrule
    $K+n\leq k <N-1$ & $(n,k-K-n)$\\
    \midrule
    $k =N-1$ & $(n,m)$\\
    \bottomrule
  \end{tabular}
  \end{center}

  \vspace{0.1cm}

  \begin{center}
  \begin{tikzpicture}[scale=1., font=\scriptsize]
    \begin{scope}[every node/.style={rectangle,thick,draw,fill=lightgray}]
      \node (0,0) at (0,0) {\scriptsize (0,0)};
      \node (1,0) at (0,-.6) {\scriptsize (1,0)};
      \node (2,0) at (0,-1.2) {\scriptsize (2,0)};
      \node (3,0) at (0,-1.8) {\scriptsize (3,0)};
      \node (0,1) at (1,0) {\scriptsize (0,1)};
      \node (1,1) at (1,-.6) {\scriptsize (1,1)};
      \node (2,1) at (1,-1.2) {\scriptsize (2,2)};
      \node (3,1) at (1,-1.8) {\scriptsize (3,2)};
      \node (0,2) at (2,0) {\scriptsize (0,2)};
      \node (1,2) at (2,-.6) {\scriptsize (1,2)};
      \node (2,2) at (2,-1.2) {\scriptsize (2,2)};
      \node (3,2) at (2,-1.8) {\scriptsize (3,2)};
      \node (0,3) at (3,0) {\scriptsize (0,3)};
      \node (1,3) at (3,-.6) {\scriptsize (1,3)};
      \node (2,3) at (3,-1.2) {\scriptsize (2,3)};
      \node (3,3) at (3,-1.8) {\scriptsize (3,3)};
      \node (0,4) at (4,0) {\scriptsize (0,4)};
      \node (1,4) at (4,-.6) {\scriptsize (1,4)};
      \node (2,4) at (4,-1.2) {\scriptsize (2,4)};
      \node (3,4) at (4,-1.8) {\scriptsize (3,4)};
    \end{scope}
    \begin{scope}[every node/.style={rectangle,thick,draw,fill=cyan}]
      \node (0,5) at (5,0) {\scriptsize (0,5)};
      \node (1,5) at (5,-.6) {\scriptsize (1,5)};
      \node (2,5) at (5,-1.2) {\scriptsize (2,5)};
      \node (3,5) at (5,-1.8) {\scriptsize (3,5)};
    \end{scope}
    \begin{scope}[every node/.style={rectangle,thick,draw,fill=cyan}]
      \node (4,0) at (0,-2.4) {\scriptsize (4,0)};
      \node (4,1) at (1,-2.4) {\scriptsize (4,1)};
      \node (4,2) at (2,-2.4) {\scriptsize (4,2)};
      \node (4,3) at (3,-2.4) {\scriptsize (4,3)};
      \node (4,4) at (4,-2.4) {\scriptsize (4,4)};
    \end{scope}
    \begin{scope}[every node/.style={rectangle,thick,draw,fill=red!50}]
      \node (4,5) at (5,-2.4) {\scriptsize (4,5)};
    \end{scope}
    \draw [->] (5.75,0) -- (6.25,0);
    \node[text width=2cm] (R0) at (7.5,0) {$R_0:\Sigma=0$};
    \draw [->] (5.75,-.6) -- (6.25,-.6);
    \node[text width=2cm] (R1) at (7.5,-.6) {$R_1:\Sigma=0$};
    \draw [->] (5.75,-1.2) -- (6.25,-1.2);
    \node[text width=2cm] (R2) at (7.5,-1.2) {$R_2:\Sigma=0$};
    \draw [->] (5.75,-1.8) -- (6.25,-1.8);
    \node[text width=2cm] (R3) at (7.5,-1.8) {$R_3:\Sigma=0$};
    \draw [->] (5.75,-2.4) -- (6.25,-2.4);
    \node[text width=2cm] (R4) at (7.5,-2.4) {$R_4:\Sigma=0$};

    \draw [->] (0,-3) -- (0,-3.5);
    \node[text width=1cm] (C0) at (0,-4) {$$C_0:$$\\$\Sigma=0$};
    \draw [->] (1,-3) -- (1,-3.5);
    \node[text width=1cm] (C1) at (1,-4) {$$C_1:$$\\$\Sigma=0$};
    \draw [->] (2,-3) -- (2,-3.5);
    \node[text width=1cm] (C2) at (2,-4) {$$C_2:$$\\$\Sigma=0$};
    \draw [->] (3,-3) -- (3,-3.5);
    \node[text width=1cm] (C3) at (3,-4) {$$C_3:$$\\$\Sigma=0$};
    \draw [->] (4,-3) -- (4,-3.5);
    \node[text width=1cm] (C4) at (4,-4) {$$C_4:$$\\$\Sigma=0$};
    \draw [->] (5,-3) -- (5,-3.5);
    \node[text width=1cm] (C5) at (5,-4) {$$C_5:$$\\$\Sigma=0$};
  \end{tikzpicture}
\end{center}

\caption{Encoding scheme: sources in gray, hyperplane parity check packets in cyan, and overall parity check packet in red.}
  \label{fig:orthotope}
\end{figure}

Informally, the encoding scheme works as follows. Place the $K=n*m$ input packets in a $n$ by $m$ rectangle (according to some ordering function which, along with $n$ and $m$ is a parameter of the problem), and add a parity check packet for each row and each column, obtained by summing all the packets in the hyperplane in question (either a row or a column). Finally, add an overall parity check packet obtained by summing all source packets, or equivalently by summing either all the row parity checks or all the column parity checks (cf Figure~\ref{fig:orthotope}). Note that if the parity check packets are placed correctly (cf Figure~\ref{fig:orthotope}), we get an $(n+1)$ by $(m+1)$ matrix where the packets in any given hyperplane (either a row or a column) sum to the zero vector. This will be the principle for decoding, but first we give a more formal definition of the encoding scheme.

These codes have linear time complexity: $O((2*K+min(m,n))*len)$ byte-xor are need for encoding and $O((2*K+m+n)*len)$ for decoding. The memory complexity for encoding and decoding is $O((m+n+1)*len)$.

\medskip

\begin{definition}[2D Rectangular Codes]
  Recall that we are working on packets of length $(len)$. Given a pair of \emph{dimensions} $(n,m)\in \mathbb{N}^*\times\mathbb{N}^*$ and a bijective ordering function $\phi:\ \mathbb{N}_{n*m}\to \mathbb{N}_n\times \mathbb{N}_m$, the \emph{two-dimensional rectangular code} with parameters $(n,m,\phi)$ is defined as follows: (where $K:=n*m$ and $N:=(n+1)*(m+1)$)
  \begin{itemize}
  \item It takes as input $\overrightarrow{X}_0,\dots,\overrightarrow{X}_{K-1}\in \{0;1\}^{(len)}$
  \item It produces as output $\overrightarrow{C}_0,\dots,\overrightarrow{C}_{N-1}\in \{0;1\}^{(len)}$, where:
    \[
    \overrightarrow{C}_i:=
    \begin{cases}
      \overrightarrow{X}_i \quad\quad\qquad\qquad\qquad\qquad\qquad\qquad\qquad\qquad \textit{if}\quad 0\leq i<K\\
      \bigoplus\limits \{\overrightarrow{X}_k : \exists y, \phi(k)=(i-K,y)\} \qquad\qquad\qquad\textit{if}\quad 0\leq i-K<n\\
      \bigoplus\limits \{\overrightarrow{X}_k : \exists x, \phi(k)=(x,i-K-n)\}  \qquad\qquad\textit{if}\quad 0\leq i-K-n<m\\
      \bigoplus\limits_{K\leq k\leq K+m}\overrightarrow{C}_k \quad\quad\qquad\qquad\qquad\qquad\qquad\qquad\textit{if}\quad i=K+n+m\\
    \end{cases}
    \]
  \end{itemize}
\end{definition}

\medskip

Note that if $n$ and $m$ are co-prime, then the ordering function $\phi$ defined by $\forall x\in \mathbb{N}_{n*m}, \phi(x)=(x \mod n,x \mod m)$ is a practical choice which makes the computation of both $\phi$ and $\phi^{-1}$  easy. Note that $\phi$ is then indeed a bijection by the Chinese remainders theorem.
Moreover, 2D rectangular codes take as input $K$ source packets, and add $M=N-K$ parity packets. In particular, they have rate $\frac{K*(len)}{N*(len)}=\frac{n}{n+1}*\frac{m}{m+1}$.

\subsubsection*{Decoding Scheme}

\begin{figure}
\begin{center}
\includegraphics[width=1.\textwidth]{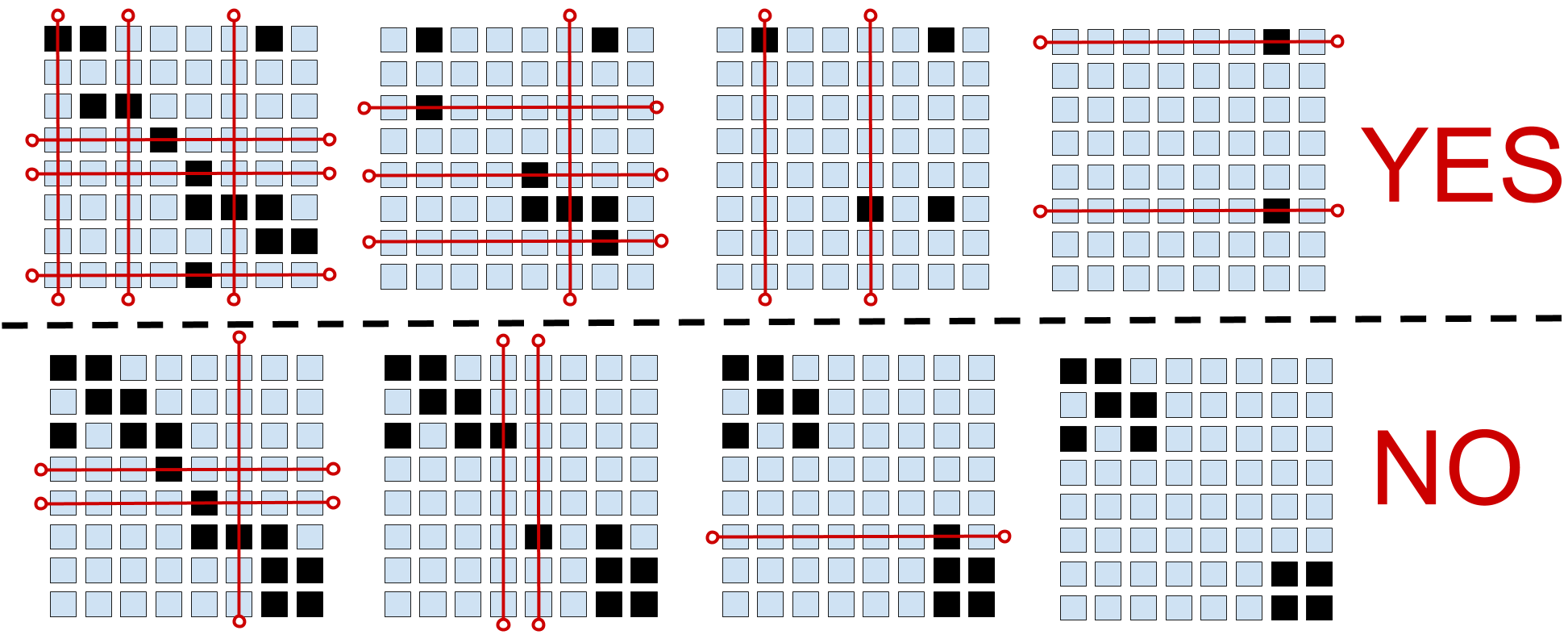}
  \caption{Visualisation of the iterative decoding algorithm of a rectangular code for the erasure channel. On top the decoding is successful. On bottom the decoding fails, due to a \textit{Stopping-set}: if the receiver wants to decode this Stopping-set, he needs at least two packets to be re-transmitted that would break these two cycles. This correspond to the \textit{Minimum Feedback Vertex Set} of this graph}. 
  \label{decoding_example}
\end{center}
\end{figure}

Recall that if the source and parity check packets are placed in a $(n+1)$ by $(m+1)$ rectangular grid in the manner described in [Figure~\ref{fig:orthotope}], it holds that the packets in any given row or column sum to zero. Therefore, if a packet is the only erroneous one in a row or column it may be repaired. This leads to a straightforward decoding algorithm:

\paragraph{Decoding Algorithm} Iteratively repair any error which is isolated in a row or column, until there are no errors left or none of them is isolated. Two examples of this iterative algorithm are given in [Figure~\ref{decoding_example}]: one leads to a successful decoding, the other to a failure, due to what is called a \textit{Stopping Set}.

\begin{figure}
\begin{center}
\includegraphics[width=.8\textwidth]{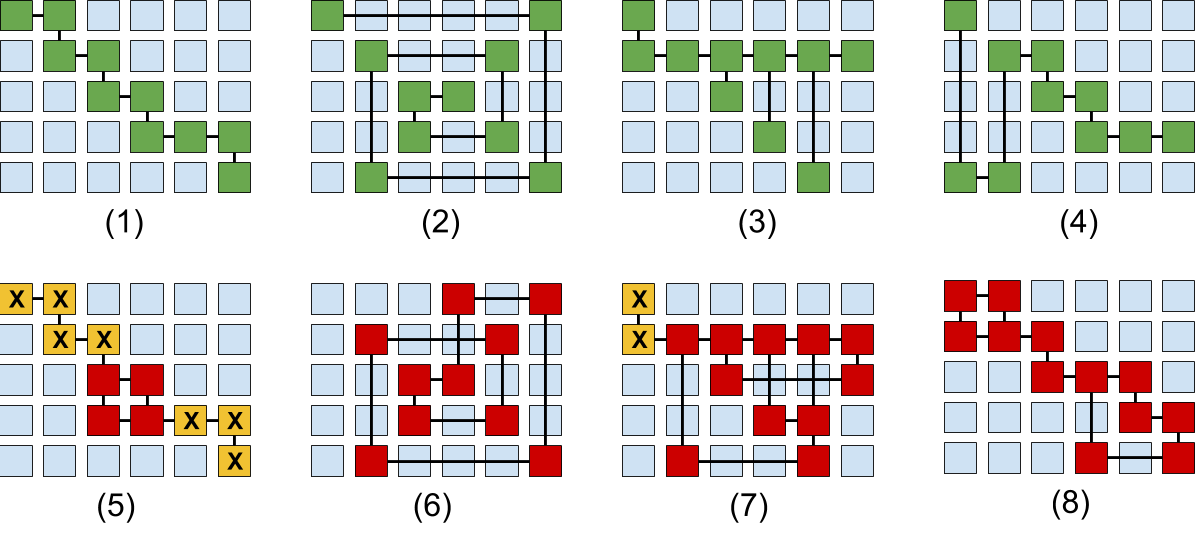}
  \caption{Error configurations: \textit{Good configurations} (1,2,3,4) without Stopping-set are fully decodable. \textit{Bad configurations} (5,6,7,8) containing Stopping-sets (in red) are not fully decodable, only the yellow packets are}.
  \label{errorConfig}
\end{center}
\end{figure}

\paragraph{Error Configurations} Define an error configuration as a set of packets erroneously transmitted. It is not always possible to repair all errors in a given configuration by inferring the erroneous packets from the correct ones without any additional information, which leads us to further characterise configurations as `Good', `Bad', or `Minimal Bad'. A \emph{good configuration} (GC) is a set of errors (possibly empty) that can be completely repaired by the decoding algorithm, a \emph{bad configuration} (BC) is a set of errors that cannot be completely repaired, and a \emph{minimal bad configuration} (MBC) is a non-empty set of errors where no error can be repaired. See [Figure~\ref{errorConfig}] for an illustration.

\subsubsection*{Feedback Repair}

Suppose that we have a bad configuration, and that we can repair erroneous packets by requesting them anew from the sender, and no longer only by inferring them from the correct ones. Assuming we have a cost function assigning to each packet a positive cost of repair, the cost of repairing a set of erroneous packets is defined as the sum of the costs of its elements. For a given configuration we define a feedback repair set as a subset of errors whose removal yields a good configuration. Note that since we require the costs of repair to be positive, the only minimum feedback repair set of a good configuration is the empty set, of cost nil.

\medskip

\noindent\textsc{MinimumFeedbackRepairSet (MinFRS):}\\
\textsc{Input:} $\mathcal{E}$ an error configuration and $w:\mathcal{E}\rightarrow \mathbb{R_+^*}$ a cost function.

\noindent\textsc{Output:} A set of errors $\mathcal{E}'\subseteq \mathcal{E}$ of minimum cost such that $\mathcal{E}\backslash\mathcal{E}'$ is a good configuration.

\medskip

\noindent\textsc{MinimumRepair (MR):}\\
\textsc{Input:} $\mathcal{E}$ an error configuration and $w:\mathcal{E}\rightarrow \mathbb{R_+^*}$ a cost function.\\
\textsc{Output:} The cost of a minimum feedback repair set.

\medskip

We use three different cost functions. The first is the \emph{all-or-none} cost function, which assigns cost one to each and every packet. It is used when requested packets must be re-transmitted in full from the sender to the receiver. The second, which is a slight variation of the first one, is the \emph{modified-all-or-none} cost function, which assigns cost one to the $\overrightarrow{C}_k$ for $k\geq K$, and cost $1+\frac{1}{N+1}$ to the $\overrightarrow{C}_k$ for $k<K$. We can use this if we prefer to request sources over parity check packets. The third is the \emph{graded} cost function where the cost of a packet is equal to the number of corrupted bits. This last function is used when we allow the receiver to request only portions of packets from the sender.

For a given cost function, two questions arise related to random error configurations.

\begin{enumerate}
\item \emph{First Question:} What is the expected amount of additional information needed to repair a random error configuration?
\item \emph{Second Question:} With what probability is no additional information needed to repair a random error configuration? In other words, with what probability is a random error configuration a good one?
\end{enumerate}

\subsection{Protocol: emission-repair cycle on data streams}
\label{sec:protocol}

In this section, we define a protocol based on iterative cycles of emissions and repairs to transmit packets data over a noisy channel. The protocol uses any black-box linear error-correcting code, but in this paper we will only use rectangular codes.

More specifically, suppose that at a given iteration we want to encode $K$ packets of data. The optimal choice of parameters $n$ and $m$ for the rectangular code is a delicate question. In order to encode the packets, we need $n\times m$ to be no smaller than $K$, and if it is greater one can pad the $n$ by $m$ rectangle by repeating some of the $K$ packets. The two-dimensional rectangular codes which minimises the rate are the square ones, i.e. those where $n=m$. So if $K$ is a perfect square, then $(\sqrt{K},\sqrt{K})$ is a good choice of parameters. If not, the problem is to achieve a satisfying trade-off between minimising the two differences $|n - m|$ and $|n*m - K|$. We do not address this problem in this paper, however, and leave open the choice of rectangular codes.

\subsubsection*{A single cycle on a single block}

We start by describing the basic building blocks of our protocol: the emission-request cycles. Suppose we have a single block of $K$ packets of input data that we want to transmit over a noisy channel. The sender encodes the input using the black-box code, and sends the yielded code packets over the channel. The receiver then determines a minimum feedback repair set,
and sends the list of the indices of the packets in the FRS to the sender using a secondary channel. This return transmission consists only of a fairly short list of integers so it is reasonable to assume that the feedback channel is error-free.

\subsubsection*{Many cycles on a single block}

Suppose we have a single block of $K$ packets of input data that we want to transmit over a noisy channel. Perform the emission-request cycle described above. The sender now holds a list of indices corresponding to a minimum FRS of the last emission. Denote by $k$ the size of the FRS. There are two possibilities: either encoding the $k$ packets requested using the black-box code yields less packets than the previous encoding of the $K$ did, or it does not. If it does, treat the $k$ packets of the FRS as the input of a new emission-request cycle. If it does not, repeat the previous emission of the $K$ packets. Note that in the latter case, the receiver determines a FRS to request based on the information received both during the previous emission and the last one. This is done in order to avoid having potentially larger and larger transmissions. Also note that if we use an $n$ by $m$ rectangular code to encode $K=n*m$ packets, then the maximum size of a minFRS over all error configurations is $k_{max}=(n+1)*(m+1)-((n+1)+(m+1)-1)=K$ and further this bound is only achieved when all of the $K$ packets are erroneous (by [Main~Theorem~\ref{expr_of_I}]), so in this particular case we can always treat the FRS as input for a new emission-request cycle.

Repeat this process while the receiver requests non-empty feedback repair sets. This process may not terminate, but if it does the receiver can retrieve the whole original block of $K$ packets by iteratively using the feedback repair sets from last to first to retrieve the first FRS, and therefore the original transmission.
\begin{figure}[!h]
  \centering
\includegraphics[width=1.0\textwidth]{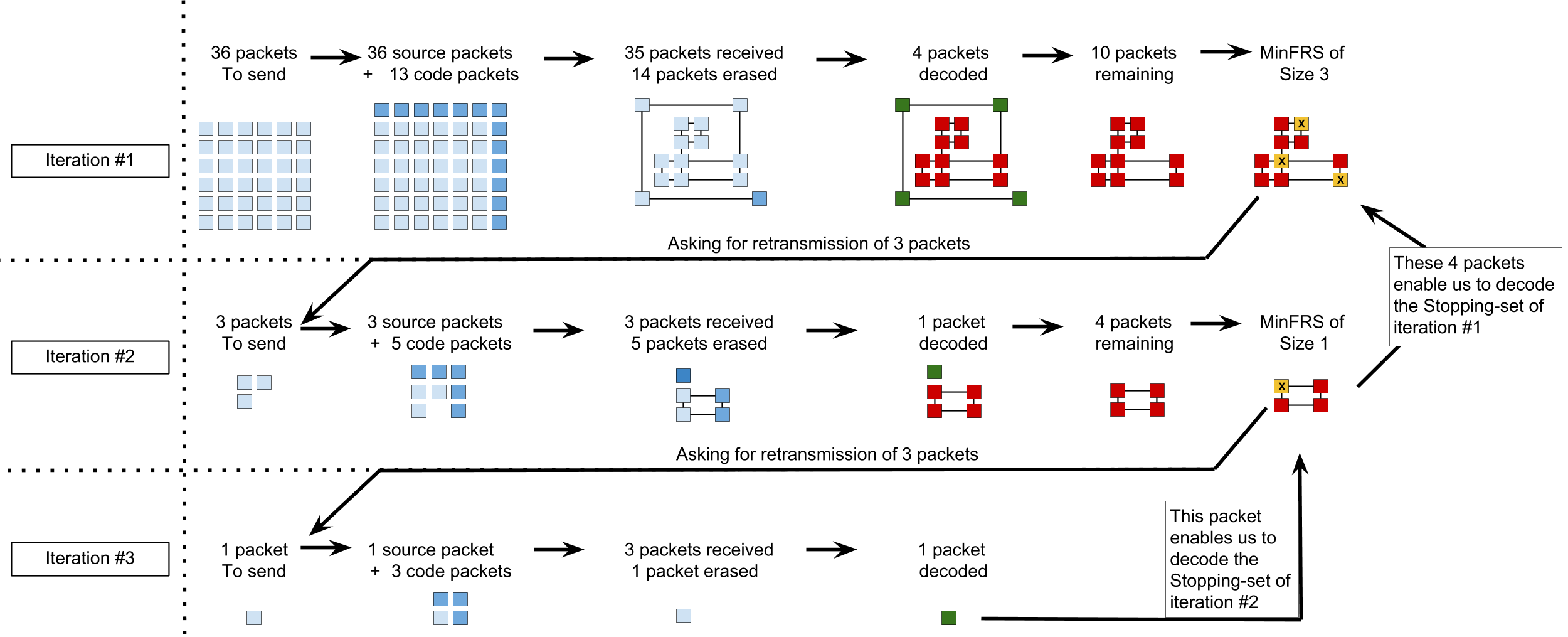}
\caption{Transmission of a single block of size $K=36$ using three iterations. At the end of iteration \#1, the decoding algorithm fails because of a Stopping-set of 10 packets: the receiver then asks for 3 packets to break this Stopping-set. At iteration \#2, The emitter sends those 3 packets plus 5 code packets, but the decoding fails once again because of a Stopping-set of 4 packets. At iteration \#3, the receiver asks for one last packet, the emitter sends 4 packets, the decoding algorithm is successful. The receiver can now decode the packets of iteration \#2, and then decode the packets of iteration \#1.}
\label{fig:protocol}
\end{figure}

\subsubsection*{Many cycles on data streams}

Suppose we have a message containing many packets of input data that we want to transmit over a noisy channel. Take a first block of $K$ packets. Perform a single emission-repair cycle, and denote by $k$ the size of the requested FRS. Add the next $(K-k)$ packets of the message to the packets of the previous FRS and perform a single emission-repair cycle. Repeat this process until there are not enough packets remaining to form a block of size $K$.

Send these remaining packets using the `Many cycles on a single block' protocol described above. If this last protocol terminates, the receiver can retrieve the whole message by iteratively using the feedback repair sets from last to first to retrieve the first FRS, and therefore the original transmission.

\subsubsection*{Memory considerations}

It is inadvisable to store too much information in the sender and receiver's respective buffers, and should forget unnecessary information. At the end of an emission-request cycle, the sender can forget all the packets in the block apart from those in the FRS that the receiver just requested, as they will never appear in any subsequent FRS. Also, in the \emph{all-or-none} repair model, once the receiver determines the list of indices of an FRS and sends it to the the sender, it can forget all the unrepaired erroneous packets it just received. Indeed, erroneous packets are repaired using only the information contained in the correct packets.


\section{Performance Analysis -- A Single Iteration}
\label{sec:analysis}

We assume that all packets are erroneous independently and with same probability $p$. Note that if $p_b$ is the bit error probability, then $p=1-(1-p_b)^{(len)}$ and so $p\approx p_b \times (len)$ for small $p_b$ and large $(len)$. The probability $p$ should be thought of as `somewhat large' (i.e. $p\in [0.2;0.8]$).

We wish to study the error configurations that we obtain by passing our message through the communication channel once using a rectangular code. Our sample space is $\mathcal{P}(\{\overrightarrow{C}_i : i\in \mathbb{N}_{N}\})$ with $\forall \mathcal{E}\subseteq \{\overrightarrow{C}_i : i\in \mathbb{N}_{N}\},\Pr(\mathcal{E})=p^{|\mathcal{E}|}(1-p)^{N-|\mathcal{E}|}$. For all $k\in \mathbb{N}_{N}$, define an indicator random variable $X_i$ associated with the packet $\overrightarrow{C}_k$ being erroneous. Let $N_e$ be the random variable whose value equals the number of erroneous packets, so that $N_e=\sum_{k=0}^{N-1}X_k$. The $(X_k)_{0\leq k<N}$ are independent and identically distributed according to a Bernouilli distribution with same parameter $p$, hence $N_e$ follows a binomial distribution of parameters $N$ and $p$. Further define indicator variables $R_i$ for $i\in \mathbb{N}_{n+1}$ and $C_j$ for $j\in \mathbb{N}_{m+1}$ associated respectively with there being an erroneous packet on the $i^{th}$ row and in the $j^{th}$ column. Let then $R$ and $C$ be the random variables denoting the total number of rows and columns respectively containing at least one erroneous packet, so that $R=\sum_{i=0}^nR_i$ and $C=\sum_{j=0}^mC_j$. Now consider the binary relation $\sim \subseteq \{\overrightarrow{C}_k, k\in\mathbb{N}_N : X_k=1\}^2$ defined by: $\overrightarrow{C}_i \sim \overrightarrow{C}_j$ ($i\neq j$) if and only if they are on the same row or in the same column when placed in the manner described in [Figure~\ref{fig:orthotope}]. Let $N_{cc}$ and $N_{nscc}$ be the random variables equal to the number of equivalence classes of respectively its reflexive, symmetric, and transitive closure $\sim^*$ and its symmetric and transitive closure $\sim^+$.
Finally, let $I$ be the random variable equal to the minimum repair cost of configuration $\mathcal{E}$. Note that $N_e$, the $C_j$, the $R_i$, $C$, $R$, $N_{cc}$, $N_{nscc}$, and $I$ are all deterministic functions of the family $(X_k)_{0\leq k<N}$, which therefore stores all of the system's randomness.

\medskip


\begin{theorem}[Main Theorem]
\emph{
  \label{expr_of_I}
  Let $\mathcal{E}$ be a random error configuration.
  If the cost function for repairs is the \emph{all-or-none} one, then there is an algorithm in $O(K)$ to find a minimum feedback repair set and furthermore:
    \begin{footnotesize}
  \begin{equation}
  I=N_e-R-C+N_{nscc}
  \label{eq:expr_of_I_Nscc}
  \end{equation}
    \end{footnotesize}
  If the cost function for repairs is arbitrary, then there is a an algorithm in $O(K+|\mathcal{E}|*log(|\mathcal{E}|))=O(K*log(K))$ to find a minimum feedback repair set.
}
\end{theorem}

\medskip

\textit{Proof.}
In the sequel, we address loopless simple undirected graphs.
For a graph $G=(V,E)$, if $V'\subseteq V$ and $E'\subseteq E$, then $G[V']$ denotes the (vertex) induced subgraph of $G$ -i.e. the graph $(V',E\cap(V')^2)$- and $G_{|E'}$ denotes the (edge) partial graph induced by $E'$ -i.e. the graph $(V,E')$.

  Assume for now that the cost function is arbitrary. Let us start by giving a sketch of the proof. First, we define a gadget that we call the \emph{coordinates' graph}, where the edges are labelled by the packets $(\overrightarrow{C}_k)_{0\leq K<N}$. We can then define the partial graph induced by an error configuration as the graph where we keep only the edges labelled by an erroneous packet. We show that for a given error configuration, feedback repair sets are exactly feedback edge sets in this partial graph. Finally, we use the fact that the partial graph induced by the complement of a feedback edge set is a forest to prove [Theorem~\ref{expr_of_I}].

\begin{figure}
\begin{center}
\includegraphics[width=.8\textwidth]{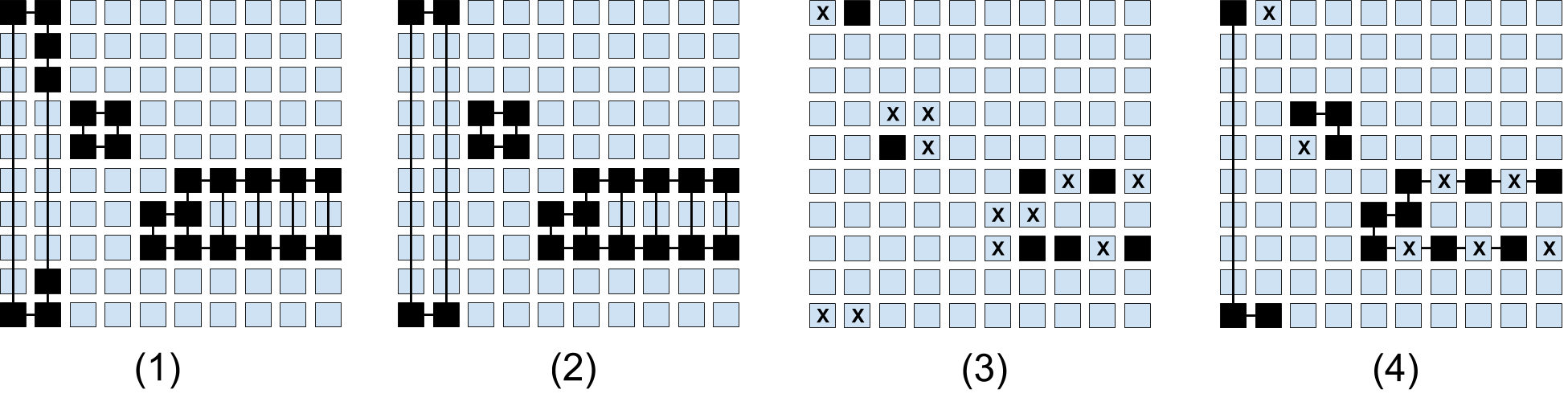}
  \caption{A \textit{bad configuration} (1) with its \textit{minimal bad configuration (2) where $n_e$=21,$R=7$,$C=10$,$N_{nscc}=3$}, a \textit{minimal FRS} (3) of size $7=21-7-10+3$, and the minimal configuration without its minimal FRS (4), \textit{containing no Stopping-set}: adding any vertex from the FRS to the remaining graph would make a Stopping-set appear.}
  \label{fvs}
\end{center}
\end{figure}

\begin{figure}
\begin{center}
\includegraphics[width=0.8\textwidth]{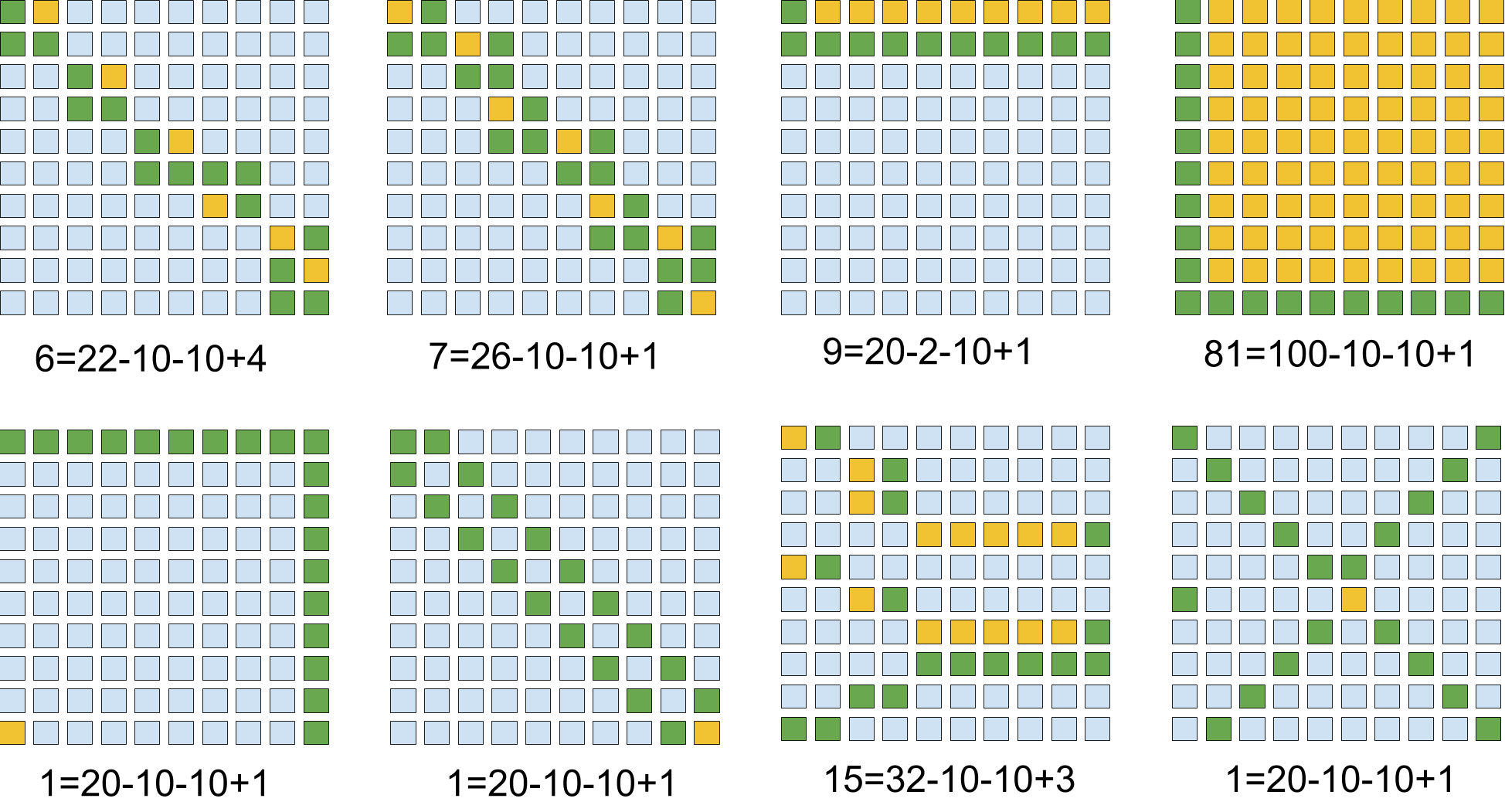}
  \caption{\textit{Minimal bad configurations} with some corresponding \textit{minimal FRS}.}
  \label{different_fvs}
\end{center}
\end{figure}

  \emph{The gadget:} Place the packets in an $(n+1)$ by $(m+1)$ grid in the manner described in [Figure~\ref{fig:orthotope}]. Define the \emph{coordinates' graph} $\mathcal{G}$ as the complete bipartite graph with $(n+1)$ vertices $(\mathcal{R}_i)_{0\leq i \leq n}$ corresponding to the rows on one side and $(m+1)$ vertices $(\mathcal{C}_j)_{0\leq i \leq m}$ corresponding to the columns on the other. For every pair $(i,j)$, label edge $(\mathcal{R}_i,\mathcal{C}_j)$ by the packet place at coordinates $(i,j)$ in the rectangular grid. Finally assign to each edge the weight given by the cost function. See [Figure~\ref{fvs}] and [Figure~\ref{different_fvs}] for different illustrations of this definition. Note that this gadget can be constructed in time $O(K+|\mathcal{E}|)=O(K)$.
  
  For a given graph, we define a \emph{feedback edge set} as a subset of edges whose removal yields a good configuration.

\medskip
  
\noindent  \textsc{MinimumFeedbackEdgeSet (MinFES):}\\
  \textsc{Input:} $G$ a weighted graph.\\
  \textsc{Output:} A (possibly empty) set of edges of minimum weight whose removal breaks all cycles.

\medskip

Having defined the gadget, we can begin the proof. Let $\mathcal{E}$ be an error configuration, and denote by $\mathcal{G}_{|\mathcal{E}}$ the partial subgraph of $\mathcal{G}$ induced by $\mathcal{E}$. The feedback repair sets of $\mathcal{E}$ are exactly the feedback edge sets of $\mathcal{G}_{|\mathcal{E}}$. Indeed, if $\mathcal{E}'\subseteq \mathcal{E}$ then $\mathcal{E}'$ is a FRS of $\mathcal{E}$ iff $\mathcal{E}\backslash \mathcal{E}'$ is a GC, i.e. $\mathcal{E}\backslash \mathcal{E}'$ is not a BC, i.e. every non-empty $\mathcal{E}''\subseteq\mathcal{E}\backslash \mathcal{E}'$ can be reduced, i.e. there is no non-empty $\mathcal{E}''\subseteq\mathcal{E}\backslash\mathcal{E}'$ containing no errors isolated in a row or column, i.e. there is no non-empty $\mathcal{E}''\subseteq\mathcal{E}\backslash\mathcal{E}'$ such that $\mathcal{G}_{|\mathcal{E}''}$ has no vertex of degree one, i.e. there is no non-empty $\mathcal{E}''\subseteq\mathcal{E}\backslash\mathcal{E}'$ such that $\mathcal{G}_{|\mathcal{E}''}$ has minimal degree two, i.e. there is no $\mathcal{E}''\subseteq\mathcal{E}\backslash\mathcal{E}'$ such that $\mathcal{G}_{|\mathcal{E}''}$ is cyclic, i.e. $\mathcal{E}\backslash\mathcal{E}'$ is acyclic, i.e. $\mathcal{E}'$ is a FES of $\mathcal{G}_{|\mathcal{E}}$. In particular, the minimum feedback repair sets of $\mathcal{E}$ are exactly the minimum feedback edge sets of $\mathcal{G}_{|\mathcal{E}}$. Therefore, the complement of a minimum FRS is a maximum forest.

In a simple, undirected, unweighted graph, a maximum forest can be found in time $O(K)$ by a \emph{depth-first search}: in each connected component, find a maximum spanning tree by arbitrarily selecting the root and then performing a depth first search, removing any backward edge encountered. Furthermore since a spanning forest has a number of edges equal to the number of vertices minus the number of connected components, we proved the part of the theorem pertaining to the \emph{all-or-none} cost function.

In a simple, undirected, weighted graph, a maximum forest can be found in time $O(K+|\mathcal{E}|*log(|\mathcal{E}|))$ with Kruskal's algorithm
(see \cite{cormen}). This proves the part of the theorem pertaining to an arbitrary cost function.
\qed

\medskip

Note that in the course of the proof, we also proved the following intermediate result which we will use again later.

\medskip

\begin{lemma}
\label{BC-characterisation}
  \emph{Let $\mathcal{E}$ be an error configuration. $\mathcal{E}$ is a good~ configuration if and only if $\mathcal{G}_{|\mathcal{E}}$ is acyclic.}
\end{lemma}

\medskip

Having established this, we can compute various indicators of the quality of our protocole. Ideally, we would like to determine the law of $I$, as well as the conditional distribution of $I$ given $N_e$. Note that this can be done in simple exponential time by exploiting the expression given in [Theorem~\ref{expr_of_I}]. Two simpler problems however are that of computing the conditional expectation instead of the conditional probability, and that of computing the conditional probability that $I=0$ given $N_e$. $\mathbb{E}(I|N_e=n_e)$ can be seen as the expected bandwidth use on the backward channel, and $\Pr(I=0|N_e=n_e)$ can be seen as the probability of successfully decoding the input data if we do not allow a backward channel. We choose to condition by the event `$N_e=n_e$' as this provides a finer-grained analysis: the performances of our protocol are averaged not over all error configurations, but over all error configurations of a given size. The law of total probability combined with [Equation~\ref{eq:law_Ne}] can however provide $\mathbb{E}(I)$ and $\Pr(I=0)$.

\subsection{Preliminary computations}

Recall that $N$, $K$, $m$, $n$, $p$ are constants while $(X_k)_k$, $N_e$, $(C_j)_j$, $C$, $(R_i)_i$, $R$, $N_{cc}$, $N_{nscc}$, $I$ are random variables.

\subsubsection*{Study of $N_e$}

Recall that $N_e$ is the random variable equal to the number of errors, and that it follows a binomial distribution of parameters $N$ and $p$. Hence, we have the following:
\begin{equation}
  \forall n_e\in \mathbb{N}_{N+1},\Pr(N_e=n_e)={N \choose n_e}*p^{n_e}*(1-p)^{N-n_e}
  \label{eq:law_Ne}
\end{equation}

\begin{equation}
\forall n_e\in \mathbb{N}_{N+1},\mathbb{E}(N_e)=p*N
\label{eq:exp_Ne}
\end{equation}

\subsubsection*{Study of $C$ and $R$ knowing $N_e$}

Recall that $C$ is the random variable equal to the number of columns containing at least one error, and that $C_j$ ($j\in \mathbb{N}_{m+1}$) is the indicator random variable corresponding to there being an error in the $j^{th}$ column.

Suppose we know that there are exactly $n_e$ errors (i.e. we condition by the event `$N_e=n_e$'), then the probabilistic setting is exactly that of throwing $n_e$ \emph{undistinguishable balls} in $(m+1)$ \emph{distinguishable bins} of maximum capacity $(n+1)$ each. Notice that the complementary event of `$C_j=0$' is `all errors occur in the $m$ other columns', i.e. in a specific area of the integral grid composed of $(N-n-1)$ points. From this, and the fact that errors occur uniformly, we get the conditional law of $C_j$ knowing $N_e$:

    \begin{footnotesize}
\begin{equation}
\forall j\in \mathbb{N}_{m+1},\Pr(C_j=0|N_e=n_e)=\frac{{N-n-1 \choose n_e}}{{N\choose n_e}}
\label{eq:law_Cj_known}
\end{equation}
    \end{footnotesize}

This in turn yields the conditional law of $C$ knowing $N_e$, as well as the conditional expectancy of $C$ knowing $N_e$ (by linearity of the expectation for instance):

\begin{equation}
\forall k\in \mathbb{N}_{m+1},\Pr(C=k|N_e=n_e)={m+1 \choose k}\left(1-\frac{{N-n-1 \choose n_e}}{{N\choose n_e}}\right)^k\left(\frac{{N-n-1 \choose n_e}}{{N\choose n_e}}\right)^{m+1-k}
\label{eq:law_C_known}
\end{equation}

\begin{equation}
\mathbb{E}(C|N_e=n_e)=\sum\limits_{j=0}^m\mathbb{E}(C_j|N_e=n_e)=(m+1)*\left(1-\frac{{N-n-1 \choose n_e}}{{N\choose n_e}}\right)
\label{eq:exp_C_known}
\end{equation}

And similarly, for the rows we obtain the following equation:


\begin{equation}
\mathbb{E}(R|N_e=n_e)=(n+1)*\left(1-\frac{{N-m-1 \choose n_e}}{{N\choose n_e}}\right)
\label{eq:exp_R_known}
\end{equation}

\subsubsection*{Study of $N_{cc}$ and $N_{nscc}$ knowing $N_e$}

Notice that for a random error configuration $\mathcal{E}$, $N_{cc}$ can be seen as the number of connected components of $\mathcal{G}_{|{\mathcal{E}}}$, and $N_{nscc}$ as the number of non-singleton connected components. It follows that $N_{cc}=N_{nscc}+(n+1-R)+(m+1-C)$. The expression of $I$ given in [Theorem~\ref{expr_of_I}] can thus be reformulated as follows:
\begin{equation}
I=N_e+N_{cc}-n-m-2
\label{eq:expr_of_I}
\end{equation}

The following theorems are due to respectively Kalugin and Saltykov, and may be found in \cite{kalugin} and \cite{saltykov} (up to the notations).

\medskip

\begin{theorem}[Kalugin ('94)]
\emph{As ${n,m,n_e\to \infty}$,
\begin{itemize}
\item if $\frac{n_e^2}{(n+1)(m+1)}\to 0$, then:
\begin{equation}
\Pr(N_{cc}=n+m+2-n_e|N_e=n_e)\to 1
\label{eq:law_Ncc_known_0}
\end{equation}
\item if $n_e=\Theta (n)$, $n_e=\Theta (m)$, and $\frac{n_e^2}{(n+1)(m+1)}\to c\in ]0;1[$, then:
\begin{equation}
\forall k\in [\![0;n_e]\!],P(N_{cc}=n+m+2-n_e+k|N_e=n_e)\to \frac{\lambda^k exp(-\lambda)}{k!},\\
where \lambda := -\frac{ln(1-c)+c}{2}
\label{eq:law_Ncc_known_1}
\end{equation}
\end{itemize}
  }
\end{theorem}

\medskip

\begin{theorem}[Saltykov ('95)]
\emph{Assume without loss of generality that $m\geq n$.
As ${(n+m)\to \infty}$, if $n_e=(n+1)*ln(n+m)+O(n)$, then:
\begin{equation}
\Pr(N_{nscc}=1|N_e=n_e)\to 1
\label{eq:law_Nscc_known}
\end{equation}
  }
\end{theorem}

\subsection{Expected bandwidth use: $\mathbb{E}(I|N_e=n_e)$}

Having done all the preliminary computations, we can now compute the first quality indicator of our protocol: the expected number of packets requested by the receiver after a single transmission containing exactly $n_e$ errors.

\medskip

\begin{theorem}[Expected bandwidth use -- known number of~errors]
\emph{Assume without loss of generality that $m\geq n$.
As ${n,m,n_e\to \infty}$,
\begin{itemize}
\item if $n_e^2=o(nm)$, then:
\begin{equation}
\mathbb{E}(I|N_e=n_e)\to 0
\label{eq:1}
\end{equation}
\item if $n_e=\Theta (n)$, $n_e=\Theta (m)$, and $\frac{n_e^2}{(n+1)(m+1)}\to x\in ]0;1[$, then:
  \begin{equation}
  \mathbb{E}(I|N_e=n_e)\to -(ln(1-x)+x)/2
\label{eq:2}
  \end{equation}
\item if $n_e=(n+1)*ln(n+m)+O(n)$, then:
\begin{equation}
  \mathbb{E}(I|N_e=n_e)=n_e+1-\frac{(m+1)*{N-n-1 \choose n_e}+(n+1)*{N-m-1 \choose n_e}}{{N\choose n_e}}+o(1)
\label{eq:3}
  \end{equation}
\end{itemize}
}
\end{theorem}

\medskip

\textit{Proof.}
~[Eq. \ref{eq:1}] is obtained by combining [Eq. \ref{eq:expr_of_I} and \ref{eq:law_Ncc_known_0}], [Eq. \ref{eq:2}] is obtained by combining [Eq. \ref{eq:expr_of_I} and \ref{eq:law_Ncc_known_1}], and [Eq. \ref{eq:3}] is obtained by combining [Eq. \ref{eq:expr_of_I_Nscc}, \ref{eq:exp_C_known}, \ref{eq:exp_R_known}, and \ref{eq:law_Nscc_known}].
\qed

\medskip

\begin{corollary}
  \emph{Assume that $n=m$. As $n\to\infty$,
\begin{itemize}
\item if $n_e=o(n)$, then:
    \begin{footnotesize}
  \begin{equation}
  \mathbb{E}(I|N_e=n_e)\to 0
  \end{equation}
    \end{footnotesize}
\item if $n_e=\Theta (n)$, and $\frac{n_e}{n}\to x\in ]0;1[$, then:
    \begin{footnotesize}
  \begin{equation}
  \mathbb{E}(I|N_e=n_e)\to \lambda(x)
  \end{equation}
    \end{footnotesize}
  where $\lambda(x)=-(ln(1-x)+x)/2\in ]0;1[$
\item if $n_e=n*ln(n)+O(n)$, then:
    \begin{footnotesize}
  \begin{equation}
  \mathbb{E}(I|N_e=n_e)=n_e+1-2(n+1)*\frac{{n^2+n \choose n_e}}{{n^2+2n+1\choose n_e}}+o(1)
  \end{equation}
    \end{footnotesize}
\end{itemize}
  }
\end{corollary}

\subsection{Probability of successfully decoding without a backward channel: $\Pr(I=0|N_e=n_e)$}

Recall from [Lemma~\ref{BC-characterisation}] that an error configuration $\mathcal{E}$ is good, i.e. $I=0$, if and only if $\mathcal{G}_{|\mathcal{E}}$ is acyclic. It follows that $\Pr(I=0|N_e=n_e)=f(n,m,n_e)/{N \choose n_e}$, where $f(n,m,n_e)$ is the number of partial subgraphs with $n_e$ edges of the labelled complete bipartite graph $K_{n,m}$. In \cite{stones}, Stones provides an exponential time algorithm to compute $f$.


\section{Experimental Results -- Many Cycles}
\label{sec:results}

\subsection{Many cycles on a single block}

In this section we present experimental results and focus on two quality indicators of the \emph{`Many cycles on a single block'} protocol: the average number of iterations and of packages transmitted before the message is received in full by the receiver. We compare these to the TCP protocol, with no forward error correction used.

\begin{figure}
\begin{center}
  \includegraphics[width=0.6\textwidth]{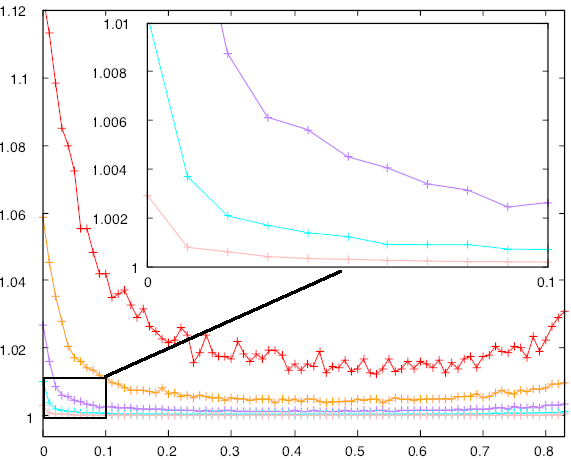}
  \caption{Average ratio between the number of packets sent ($e_K$) until transmission is over (our solution divided by the tcp solution) for $K = 256$ (red), $K = 1024$ (orange), $K = 4096$ (purple), $K = 16384$ (cyan), $K = 65356$ (pink) as functions of the package error rate $p$.}
  \label{efficiency}
\end{center}
\end{figure}

Define the \emph{efficiency} $e_K$ of the \emph{`many cycles on a single block of size $K$'} protocol as the number of packages sent by the sender over all iterations divided by the number of packages transmitted (i.e. $K$). Note that $e_K \geq 1$. We plot $e_K$ for $K\in \{256,1024,4096,16385,65536\}$ in [Figure~\ref{efficiency}] as functions of $p$. We would like this value to be the nearest possible to 1. [Figure~\ref{efficiency}] shows that for $p = 0$, every packet is received, so we have $e_K = \frac{(m+1)*(n+1)}{K} \longrightarrow 1 + 2\sqrt{K}$, whereas for greater values of $p$, $e_K \longrightarrow 1$. To reduce this value, we could consider another strategy: we could send only the $K$ source packets at the first iteration, and send some code packets at the second iteration, if needed. This way, we would decrease the average number of packets sent, but we would increase the average number of iterations required, which is what we try to avoid.
\begin{figure}
\begin{center}
  \includegraphics[width=0.6\textwidth]{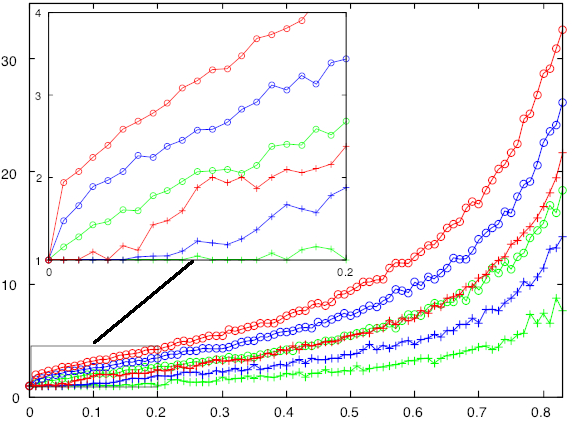}
  \caption{Average number of iterations $(i_K)$ until transmission is over for $K = 16$ (red), $K = 64$ (green), $K = 256$ (blue) as functions of the package error rate $p$. Lines with circles are for tcp solution, and lines with cross are for our solution.}
\label{iterations}
\end{center}
\end{figure}

\begin{figure}
\begin{center}
  \includegraphics[width=0.6\textwidth]{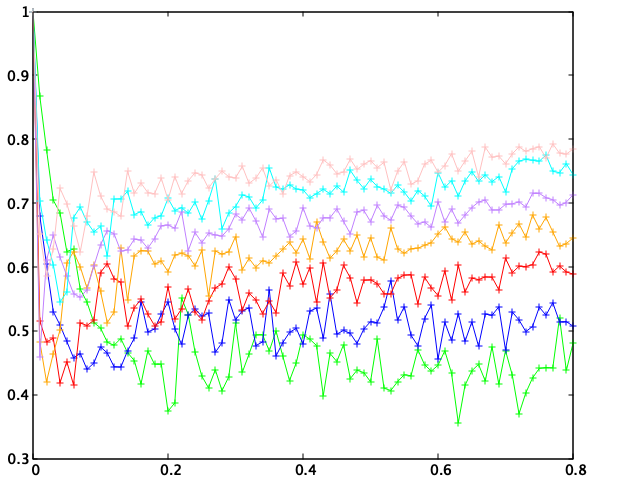}
  \caption{Average gain in terms of iterations $(i_K)$ using our protocol compared to tcp protocol until transmission is over for $K = 16$ (green), $K = 64$ (blue), $K = 256$ (red), $K = 1024$ (orange), $K = 4096$ (purple), $K = 16384$ (cyan), $K = 65536$ (pink) as functions of the package error rate $p$.}
\label{gain}
\end{center}
\end{figure}

    Define $i_K$ as the average number of iterations before the \emph{`many cycles on a single block of size $K$'} protocol terminates. We plot $i_K$ for $K\in \{16,64,256\}$ in [Figure~\ref{iterations}]. For example, sending 256 packets with TCP protocol over a channel with an erasure probability $p = 0.3$ will require about 6 iterations, whereas it requires approximately 3 iterations with our protocol. [Figure~\ref{gain}] shows the average gain in terms of number of iterations between our protocol and the TCP protocol. For $K=16$, our protocol requires approximately $60\%$ less iterations than TCP, for $K = 256$ it requires $50\%$ less iterations, and for $K = 65536$, it requires approximately $25\%$ less iterations. As for the H-ARQ protocol, if $p$ is great enough for the code to be useful (i.e: we lose at least one packet at the first iteration), we can observe an improvement in terms of iterations.

    These experimental results were obtained as follows. For each value of $K$, the protocol was run $10^5$ times, each time with an erasure probability $p$ taken uniformly at random in the interval $]0;1[$. Recall from [Section~\ref{sec:protocol}], that the choice of parameters $n$ and $m$ at a given iteration is problematic. These results were obtained by `padding with zeros': $n$ and $m$ are chosen equal and minimal such that $n^2\geq K$. $(n^2-K)$ empty packages (ie. $(0,\dots,0)$) are added to complete the square code, but are never sent.

\section{Conclusion}
\label{sec:conclusion}

The heart of the protocol presented in this paper is the minimality of feedback information. What we require from the ECC is that there exists an \emph{efficient} algorithm to determine a minFRS. Potential candidates for the ECC could be higher dimensional rectangular codes, however no minimum feedback algorithm is known to us. The error-detecting scheme can be changed too: we can use a minimal distance separable ECC instead of the UDP protocol to generate authenticated packets, which then have a chance of `repairing themselves'.

As it stands, using two dimensional rectangular codes, our protocol presents two potential use-cases directly linked to the minimality of the feedback information requests:
\begin{itemize}
\item \emph{Minimally bidirectional channels:} Data diodes are used in guaranteeing information security, but strictly unidirectional channels cannot achieve zero-error transmissions. Our hybrid protocol however can, and limits bidirectionality to what is necessary. Further, all computations (encoding, feedback information determination, and decoding) are sufficiently light to be performed by hardware inside the data diode.
\item \emph{Reducing round-trip time:} Our protocol can be seen as an `improved TCP' since the addition of an error-correcting code reduces transmission time.
\end{itemize}



\bibliography{references}
\bibliographystyle{plain}

\end{document}